\documentclass[prc,aps,groupedaddress,superscriptaddress,twocolumn,nofootinbib,showpacs,showkeys,floatfix,a4paper,10pt,amsmath,amssymb]{revtex4-1}
\usepackage{graphicx,here}
\usepackage{dcolumn}
\usepackage{bm}
\usepackage{hyperref}
\hypersetup{colorlinks=True,urlcolor=blue,linkcolor=blue,citecolor=blue,filecolor=black}

\begin{document}

\title{Microscopic description of quadrupole-hexadecapole coupling 
in radium, thorium, uranium and plutonium isotopes
with the Gogny energy density functional}

\author{R. Rodr\'{\i}guez-Guzm\'an}
\email{guzman.rodriguez@nu.edu.kz}
\affiliation{Department of Physics, Nazarbayev 
University, 53 Kabanbay Batyr Ave., Astana 010000, Kazakhstan}

\author{L. M. Robledo}
\email{luis.robledo@uam.es}
\affiliation{%
Center for Computational Simulation, Universidad Polit\'ecnica de 
Madrid, Campus Montegancedo, 28660 Boadilla del Monte, Madrid, Spain
}%
\affiliation{Departamento  de F\'{\i}sica Te\'orica and CIAFF, 
Universidad Aut\'onoma de Madrid, 28049-Madrid, Spain}

\date{\today}

\begin{abstract} 
The emergence and stability of static hexadecapole deformations as well 
as the impact in the development of dynamic deformation due to collective
motion considering quadrupole-hexadecapole coupling 
are studied for a selected set of radium, thorium, uranium and 
plutonium isotopes, using the Gogny Hartree-Fock-Bogoliubov and 
Generator Coordinate Method frameworks. Sizable hexadecapole 
deformations are found to play a significant role in the ground and 
excited states of nuclei in the neighborhood of  $^{238}$U. For each of 
the studied isotopic chains, it is shown that a region with small 
negative hexadecapole deformation, just below the neutron magic number 
$N =184$, remains stable once zero-point quadrupole-hexadecapole 
fluctuations are taken into account. A transition is predicted, with 
increasing mass number, from a regime in which the quadrupole and 
hexadecapole degrees of freedom are interwoven  to  a regime in which 
they are decoupled, accompanied by an enhanced shape coexistence in the 
more neutron-rich sectors of the isotopic  chains. It is also shown, 
that quadrupole-hexadecapole configuration mixing brings a nontrivial  
additional correlation energy gain comparable to the quadrupole 
correlation energy itself.
\end{abstract}


\maketitle{}

%
%
%

\section{Introduction.}
A significant portion of our present understanding of low-energy 
nuclear structure hinges on the concept of intrinsic deformation, 
rooted in the spontaneous symmetry breaking mechanism of (product) 
mean-field Hartree-Fock-Bogoliubov (HFB) states $| \varphi \rangle$. 
Nuclear spectra are well known to exhibit fingerprints of several types  
of shape degrees of freedom  associated to intrinsic deformations 
\cite{rs}. Therefore, a better understanding of those degrees of 
freedom and their interplay, still remains a central issue in nuclear 
structure  nowadays.

Among the  multipole moments of the nuclear density, the axial 
quadrupole moment $Q_{20}$, associated with the deformation parameter 
$\beta_{2}$, is the most relevant and driving one  from the energetic 
point of view. However,  other  deformations may play a role in 
specific regions of the nuclear chart. For example, triaxial and/or 
$\gamma$-soft ground states have been predicted in certain regions 
\cite{triaxial-example-1,triaxial-example-2,triaxial-example-3}, and 
the impact of $(\beta_{2},\gamma)$-deformations on the spectroscopic 
properties of atomic nuclei, has  been studied using a variety of 
theoretical approaches 
\cite{Others-example-1,Others-example-2,Others-example-3,Others-example-4,Others-example-5,triaxal-example-8,triaxal-example-9,triaxal-example-10,triaxal-example-11}. 
One should also keep in mind, that the heights of the static inner 
fission barriers in heavy and super-heavy nuclei are reduced when the 
$\gamma$ degree of freedom is included in the calculations 
\cite{triaxial-example-4,triaxial-example-5,triaxial-example-6,triaxial-example-7}. 

In regions with proton and/or neutron numbers around 
34, 56, 88 and 134 -where the coupling between
intruder $(N + 1, l + 3, j + 3)$ and normal-parity $(N, l, j)$
states is more pronounced- the spatial reflection symmetry is spontaneously 
broken \cite{q3-review} and octupole-deformed ground 
states are favored energetically. Octupole 
deformation has a strong impact
in the outer sectors of the fission paths in
heavy and super-heavy nuclei 
\cite{triaxial-example-6,outer-barriers-1,outer-barriers-2} and plays a 
key role as collective variable in  cluster 
radioactivity \cite{cluster}. This kind of 
deformation has received 
close scrutiny 
using  approaches such as the macroscopic-microscopic 
(Mac-Mic) model \cite{Mac-Mic-octupole-1,Mac-Mic-octupole-2}, the (mapped) Interacting Boson Model (IBM) 
\cite{mapped-IBM-oct-1,mapped-IBM-oct-2,mapped-IBM-oct-3,mapped-IBM-oct-4,mapped-IBM-oct-5,mapped-IBM-oct-6,mapped-IBM-oct-7}
as well as microscopic approaches, both at the mean-field level and beyond 
\cite{micro-oct-studies-1,micro-oct-studies-2,micro-oct-studies-3,micro-oct-studies-4,micro-oct-studies-5,micro-oct-studies-6,micro-oct-studies-7,micro-oct-studies-8,micro-oct-studies-9,micro-oct-studies-10,micro-oct-studies-11}.

In the case of the Gogny energy density functional (EDF) \cite{gogny}
previous microscopic studies 
\cite{2DGCM-q2q3-Gogny-1,2DGCM-q2q3-Gogny-2,2DGCM-q2q3-Gogny-3,2DGCM-q2q3-Gogny-4,2DGCM-q2q3-Gogny-5} have considered
the quadrupole-octupole coupling in different regions of the 
nuclear chart via  
configuration mixing calculations within the framework of the 
two-dimensional (2D)
Generator Coordinate Method (GCM) \cite{rs}. Those 
studies provided a better understanding of the emergence of static octupole deformation
effects  within the Gogny-HFB framework and the 
stability of such effects once  dynamical correlations, not explicitly 
taken into account at the mean-field level, are considered. It has also been found 
that in most cases, the 
quadrupole-octupole coupling is rather weak, i.e., to 
a large extent, the octupole dynamics can be accounted for in one-dimensional (1D) GCM 
calculations with the octupole 
deformation as single generating coordinate \cite{2DGCM-q2q3-Gogny-1,2DGCM-q2q3-Gogny-2,2DGCM-q2q3-Gogny-3,2DGCM-q2q3-Gogny-4,2DGCM-q2q3-Gogny-5}.

In comparison with the quadrupole and octupole cases, hexadecapole deformation
has received less detailed attention
\cite{triaxial-example-6,hexa-prvious-1,hexa-prvious-2,hexa-prvious-3,hexa-prvious-4}.
In the case of rare-earth nuclei, this type of deformation can affect the values of the 
neutrino-less double $\beta$ decay
matrix elements \cite{NL-bb-paper}. For a recent analysis of the impact of 
hexadecapole deformations on the collective spectra of
axially deformed nuclei within the (mapped) IBM approach, the 
reader is referred to Ref.~\cite{Nomura-Lotina-b4}.
In fission calculations, hexadecapole 
deformations play an important role  
for near-to-scission 
configurations  \cite{A.Zdeb-scission}. From the 
experimental point of view, hexadecapole $K^{\pi}=4^{+}$ vibrational
bands have been identified  \cite{exp-b4-1,exp-b4-2,exp-b4-3}
in previous studies. Both quadrupole  and hexadecapole deformations  have been  
inferred from   inelastic proton scattering experiments in inverse kinematics on 
$^{74}$Kr and $^{76}$Kr \cite{scattering-b2b4}.

In a recent large-scale HFB survey for even-even nuclei 
\cite{large-scale-b4}, regions of the nuclear chart where static 
hexadecapole $\beta_{4}$ deformations are favored energetically, have 
been identified. To this end, calculations were carried out using the 
parametrization D1S \cite{gogny} and D1M$^{*}$ \cite{gogny-d1mstar} of 
the Gogny-EDF. These Gogny-HFB calculations were followed by an account 
of the dynamical $(\beta_{2},\beta_{4})$-coupling in Sm and Gd isotopes 
within the 2D-GCM framework. A key lesson extracted from those 2D-GCM 
calculations is that, at variance with the rather weak 
quadrupole-octupole coupling found in similar calculations 
\cite{2DGCM-q2q3-Gogny-1,2DGCM-q2q3-Gogny-2,2DGCM-q2q3-Gogny-3,2DGCM-q2q3-Gogny-4,2DGCM-q2q3-Gogny-5}, 
the quadrupole and hexadecapole degrees of freedom are strongly coupled 
and, therefore, full-fledged 2D-GCM calculations must be carried out. 
Furthermore, at least for some of the considered Sm and Gd nuclei, the 
inclusion of the hexadecapole degree of freedom in the ground state 
dynamics leads to an additional correlation energy around 500-600 keV, 
which is similar in size to typical quadrupole correlation energies.

The results  mentioned above suggest that it is timely and necessary to 
examine the nontrivial beyond-mean-field 
$(\beta_{2},\beta_{4})$-coupling in regions other than the Sm and Gd 
isotopic chains, starting from modern  EDFs with reasonable predictive 
power all over the nuclear chart. This is our main goal in this work, 
devoted to the study of hexadecapole deformation effects and their 
coupling to the quadrupole degree of freedom in several isotopic chains 
(Ra, Th, U and Pu) in the actinide region. Our study is also motivated 
by a recent analysis of hydrodynamic simulations of the elliptic flow 
in heavy ion collisions involving the creation of a quark-gluon plasma 
at the BNL Relativistic Heavy Ion Collider (RHIC), which demonstrated 
the key role of hexadecapole deformation to  restore the agreement 
between the simulations and the available data for collisions of 
$^{238}$U \cite{b4-RIHC}.

In this paper, we address the emergence and stability of mean-field 
hexadecapole deformation effects as well as the impact of the dynamical  
$(\beta_{2},\beta_{4})$-coupling in the long isotopic chains 
$^{232-268}$Ra (Z=88), $^{232-268}$Th (Z=90), $^{232-268}$U (Z=92) and  
$^{232-268}$Pu (Z=94). The Z values lie around the region where the 
largest $\beta_4$ values are expected in the actinide region 
\cite{large-scale-b4}. The reason driving our choice of the mass range 
$232 \le A \le 268$ is, first, to examine both static and dynamic 
$\beta_{4}$ deformation effects around $^{238}$U \cite{b4-RIHC}. 
Second, the considered mass range extends up to neutron-rich isotopes 
below the magic neutron number $N = 184$ \cite{triaxial-example-6}. It 
is precisely in this region where static $\beta_{4} < 0$ values are 
obtained in previous Gogny-HFB calculations \cite{large-scale-b4}, in 
agreement with the predictions of the polar-gap model 
\cite{polar-gap-1,polar-gap-2}. 

To the best of our knowledge, a detailed dynamical analysis of 
$\beta_{4} < 0$ deformation has not yet been presented in the 
literature. On the other hand, the predicted region with  small 
$\beta_{4} < 0$ values requires  a dynamical analysis  to asses the 
survival of those weak hexadecapole deformation effects from a 
beyond-men-field perspective.  Furthermore, such a dynamical analysis 
is also required  since the structural evolution of the mean-field 
potential energy surfaces (MFPESs) in the considered isotopic chains 
(see, Sec.\ref{MF_RESULTS}) reveals that, as we move to more 
neutron-rich sectors, there is a pronounced competition between 
low-lying configurations (minima) based on different intrinsic 
(quadrupole and/or hexadecapole) shapes.

As in previous studies \cite{2DGCM-q2q3-Gogny-1,2DGCM-q2q3-Gogny-2, 
2DGCM-q2q3-Gogny-3,2DGCM-q2q3-Gogny-4,2DGCM-q2q3-Gogny-5,large-scale-b4}, 
we have resorted to a theoretical framework consisting of two steps. 
First, we have performed HFB calculations with constrains \cite{rs} on 
the axially symmetric quadrupole $\hat{Q}_{20}$ and hexadecapole 
$\hat{Q}_{40}$ operators. For each of the studied nuclei, the HFB 
calculations provide the energies $E_{HFB}(\beta_{2},\beta_{4})$ and 
the corresponding intrinsic states $| \varphi(\beta_{2},\beta_{4}) 
\rangle$ labeled by the quadrupole $\beta_{2}$ and hexadecapole  
$\beta_{4}$ deformation parameters [see, Eq.(\ref{def-betas}) below]. 
As a second step, the HFB states $| \varphi(\beta_{2},\beta_{4}) 
\rangle$ are used as a basis in a 2D-GCM ansatz. Via the solution of 
the corresponding Griffin-Hill-Wheeler (GHW) equation \cite{rs}, we 
study the stability of static hexadecapole deformation effects once 
zero-point quantum fluctuations in the generating 
$(\beta_{2},\beta_{4})$-coordinates are taken into account.

Both at the HFB and 2D-GCM levels, we have resorted to the 
parametrization D1S of the Gogny-EDF \cite{gogny}. Such a 
parametrization has already been shown to provide a reasonable 
description of several properties, both at the mean-field level and 
beyond, all over the nuclear chart \cite{review-RRR-2019}. The 
Gogny-D1S parametrization has also been successfully employed, as a 
reference, in a previous large-scale HFB survey of hexadecapole 
properties \cite{large-scale-b4}. We have also carried out calculations 
with the parametrization D1M$^{*}$ \cite{gogny-d1mstar} and D1M 
\cite{gogny-d1m}. However, as in previous studies (see, for example, 
\cite{2DGCM-q2q3-Gogny-1,2DGCM-q2q3-Gogny-2, 
2DGCM-q2q3-Gogny-3,2DGCM-q2q3-Gogny-4,2DGCM-q2q3-Gogny-5,large-scale-b4} 
and references therein), we have found a very mild dependence of the 
results on the underlying parametrization of the Gogny-EDF and, 
therefore, we will not dwell on these details in the present study.

Considering hexadecapole triaxial degrees of freedom would be a natural
extension to the present approach as it would give access to study 
the properties of the $K^{\pi}=4^{+}$ vibrational
bands recently identified  \cite{exp-b4-1,exp-b4-2,exp-b4-3}. However,
the extra complexity of both going triaxial and dealing at the same 
time with three collective
coordinates makes the calculations extremely expensive.

The paper is organized as follows. The employed theoretical framework 
is briefly outlined in Sec.~\ref{Theory}. In particular, the Gogny HFB 
and 2D-GCM approaches are briefly sketched in 
Secs.~\ref{Theory-Gogny-MF} and \ref{Theory-Gogny-2D-GCM}. For a more 
detailed account the reader is referred to 
Refs.~\cite{2DGCM-q2q3-Gogny-1,large-scale-b4,review-RRR-2019}. The 
results of the calculations are discussed in Sec.~\ref{results}. First, 
in Sec.~\ref{MF_RESULTS} attention is  paid to the structural evolution 
of the MFPESs leading to the emergence of (static)  positive and/or 
negative hexadecapole deformation effects in the isotopic chains 
$^{232-268}$Ra, $^{232-268}$Th, $^{232-268}$U and $^{232-268}$Pu. The 
2D-GCM results obtained for those nuclei are  discussed in 
Sec.~\ref{GCM_RESULTS}. Here, attention is paid to the stability of 
static $\beta_{4}$ deformation effects and the impact of the 
beyond-mean-field  $(\beta_{2},\beta_{4})$-coupling via the analysis of 
quantities such as collective wave functions, dynamical deformation 
values and correlation energies. Results obtained with 1D-GCM 
calculations,  will also be discussed in this section. In the same 
section, we examine the interleaving of the quadrupole and hexadecapole 
degrees of freedom in the excited 2D-GCM states. Finally, 
Sec.~\ref{conclusions} is devoted to the concluding remarks.

%
%

\section{Theoretical framework}
\label{Theory}


To study the interplay between the quadrupole and hexadecapole degrees 
of freedom, both at the (static) HFB and dynamical (2D-GCM) levels, we 
have resorted in this work to the  HFB+2D-GCM scheme 
\cite{2DGCM-q2q3-Gogny-1,2DGCM-q2q3-Gogny-2, 
2DGCM-q2q3-Gogny-3,2DGCM-q2q3-Gogny-4,2DGCM-q2q3-Gogny-5,large-scale-b4}, 
based on the D1S parametrization of the Gogny force \cite{gogny}. The 
key ingredients of the HFB approach are sketched in 
Sec.~\ref{Theory-Gogny-MF}, while the 2D-GCM scheme is briefly outlined 
in Sec.~\ref{Theory-Gogny-2D-GCM}.


\subsection{The Gogny HFB framework}
\label{Theory-Gogny-MF}


Our starting theoretical tool is the constrained HFB method \cite{rs}. 
Aside from the usual mean-field constrains on the proton and neutron 
numbers \cite{rs}, the HFB equation has been solved with constrains on 
the axially symmetric quadrupole $\hat{Q}_{20}$ and hexadecapole 
$\hat{Q}_{40}$ operators. The  quadrupole $Q_{2 0}$ and hexadecapole 
$Q_{4 0}$ moments are obtained as average values of the corresponding 
multipole operators $Q_{\lambda 0} = \langle \varphi | \hat{Q}_{\lambda 
0} | \varphi \rangle$ ($\lambda = 2,4)$ in the HFB states $| \varphi 
\rangle$. We have parameterized the moments $Q_{\lambda 0}$ in terms of 
the deformation parameters $\beta_{\lambda}$  \cite{defb2b4theory}
\begin{equation} \label{def-betas}
Q_{\lambda 0} = \frac{3 R_{0}^{\lambda} A}{\sqrt{4 \pi (2\lambda +1)}} \beta_{\lambda}
\end{equation}
with $R_{0}=1.2A^{1/3}$ and $A$ the mass number. Thus, the label 
$\vec{\beta}= (\beta_{2},\beta_{4})$ is attached to the HFB energies 
$E_{HFB}(\vec{\beta})$ and intrinsic states $| \varphi (\vec{\beta}) 
\rangle$. For the solution of the constrained HFB equations, an 
approximate second order gradient method has been employed \cite{SOGM}. 

Axial symmetry has been kept as a self-consistent symmetry in the 
calculation as triaxial shapes are not expected to play a relevant role 
in the ground state properties. The HFB quasiparticle operators 
\cite{rs} have been expanded in a large  axially symmetric harmonic 
oscillator (HO) basis  containing 17 major shells with the oscillator 
lengths $b_{\perp}=b_{z}= b_{0}=1.01 A^{1/6}$. In the HFB calculations 
all the states $| \varphi (\vec{\beta}) \rangle$ are characterized by 
the same value of the oscillator length $b_{0}$ in order to avoid 
technical difficulties in the application of the extended Wick theorem 
\cite{EWT-1,EWT-2} to the calculation of GCM kernels. To test the
convergence with basis size, calculations with 19 major shells have
been performed in a handful of cases. The results indicate that 
apart from a slight increase in the total binding energy, other
observables like excitation energies and transition probabilities
remain essentially unaltered.
 
As  a result of the mean-field approach, we have obtained a set of  
intrinsic states and  energies in a $(\beta_{2},\beta_{4})$-mesh  with 
the step sizes  $\delta \beta_{2}= \delta \beta_{4}= 0.01$. Given the 
possibility to find in the same nucleus coexisting prolate and oblate 
structures as well as very flat minima in the hexadecapole direction a 
large deformation domain has been considered in the HFB calculations
\begin{equation}
\label{2D-mesh}
{\cal{D}}_{\vec{\beta}}= \Big\{\beta_{2} \in [-0.8,1.2]; \beta_{4} \in [-0.8,1.2] \Big\}
\end{equation}
We find that the given domain is enough to account for basic properties 
in both the ground and excited states of the studied nuclei at the 
2D-GCM level.
 

\subsection{The Gogny 2D-GCM framework}
\label{Theory-Gogny-2D-GCM}


Once the HFB states $| \varphi (\vec{\beta}) \rangle$ have been 
obtained in the deformation domain ${\cal{D}}_{\vec{\beta}}$, they are 
employed as basis for the 2D-GCM ansatz
\begin{equation} \label{GCM-WF}
| {\Psi}_{2D-GCM}^{\sigma} \rangle = \int_{{\cal{D}}_{\vec{\beta}}} d\vec{\beta} f^{\sigma} (\vec{\beta}) | {\varphi} (\vec{\beta}) \rangle
\end{equation}
where the index $\sigma$ numbers the ground $\sigma = 1$ and excited 
$\sigma = 2,3, \ldots$ GCM solutions. We have also introduced the 
shorthand notation $\vec{\beta}=(\beta_{2},\beta_{4})$ in the above 
equation. In the numerical applications the integrals in 
Eq.(\ref{GCM-WF}) have been discretised in the domain 
${\cal{D}}_{\vec{\beta}}$ with the step size given above.
 
A more general GCM ansatz than the one in Eq.(\ref{GCM-WF}) would be 
required to consider the coupling to additional degrees of freedom 
\cite{other-degrees-1}. Nevertheless, for obvious technical reasons 
-such as the large number of HO shells used and the large deformation 
domain employed in the calculations- such a task is out of the scope of 
this paper, where we will restrict to the use of the $\vec{\beta}= 
(\beta_{2},\beta_{4})$ generating coordinates.

The amplitudes $f^{\sigma} (\vec{\beta})$ in the GCM ansatz of Eq.(\ref{GCM-WF}) are determined 
dynamically via the solution of the 
GHW equation  \cite{rs,2DGCM-q2q3-Gogny-1,2DGCM-q2q3-Gogny-2,
2DGCM-q2q3-Gogny-3,2DGCM-q2q3-Gogny-4,2DGCM-q2q3-Gogny-5,large-scale-b4}
\begin{equation} \label{GHE-eq}
\int_{{\cal{D}}_{\vec{\beta}_{2}}} d\vec{\beta}_{2}
\Big\{{\cal{H}}(\vec{\beta}_{1},\vec{\beta}_{2}) - 
E^{\sigma} {\cal{N}}(\vec{\beta}_{1},\vec{\beta}_{2})\Big\} f^{\sigma} (\vec{\beta}_{2}) 
=0
\end{equation}
written in terms of non-diagonal norm 
\begin{equation} \label{NKGCM}
{\cal{N}}(\vec{\beta}_{1},\vec{\beta}_{2})= \langle \varphi(\vec{\beta}_{1})| \varphi(\vec{\beta}_{2}) \rangle
\end{equation}
and Hamiltonian 
\begin{equation} \label{HKGCM}
{\cal{H}}(\vec{\beta}_{1},\vec{\beta}_{2})= \langle \varphi(\vec{\beta}_{1})| \hat{H}| \varphi(\vec{\beta}_{2}) \rangle
\end{equation}
kernels. In the evaluation of the density-dependent contribution to the 
Hamiltonian kernel ${\cal{H}}(\vec{\beta}_{1},\vec{\beta}_{2})$, we 
have considered the mixed-density prescription \cite{Sheikh21}. 
Perturbative first-order corrections in both the mean value of the 
proton and neutron numbers have also been included in the calculations 
\cite{EWT-2,Sheikh21,2DGCM-q2q3-Gogny-1,2DGCM-q2q3-Gogny-2, 
2DGCM-q2q3-Gogny-3,2DGCM-q2q3-Gogny-4,2DGCM-q2q3-Gogny-5}.

\begin{figure*}
\includegraphics[width=1.0\textwidth]{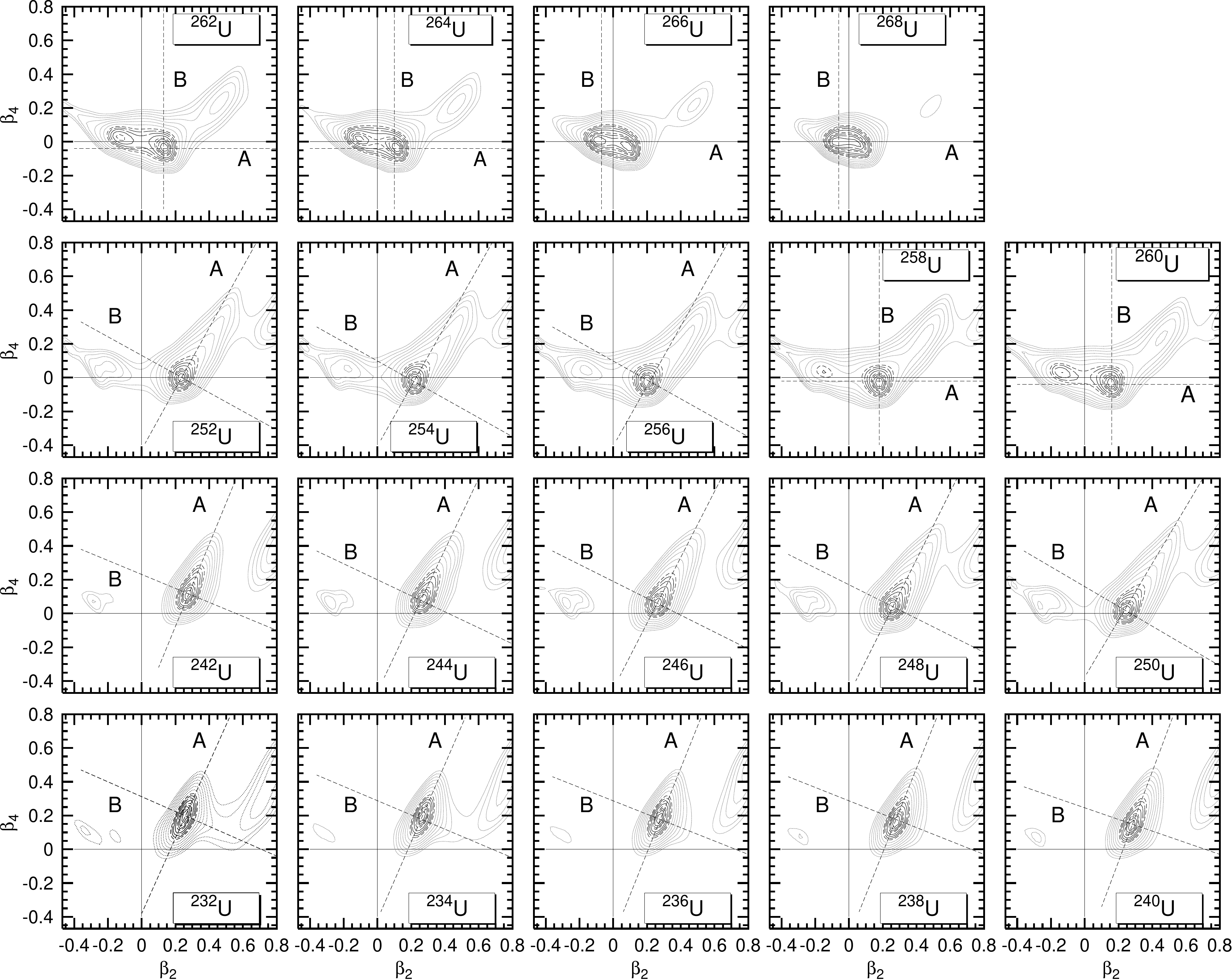}
\caption{ MFPESs computed with the Gogny-D1S EDF for the 
isotopes $^{232-268}$U. Contour lines extend from 0.25 MeV up 
to 1 MeV above the ground state energy in steps of 0.25 MeV in the 
ascending sequence full, long-dashed, medium-dashed and short-dashed. 
The next contours following the same sequence correspond to
energies from 1.5 MeV up to 3 MeV above the 
ground state in steps of 0.5 MeV. From there on, 
dotted contour lines are drawn in steps of 1 MeV. For each nucleus, the two 
perpendicular dotted lines A and B are drawn along the principal axes 
of the parabola that approximates the HFB energy around the absolute 
minimum of the MFPES. A vertical full line is drawn to signal the $\beta_{2}=0$
line whereas a full horizontal line is drawn to signal the $\beta_{4}=0$ line.  For more details, see the main text.
}
\label{mean-field-surfaces-U-1} 
\end{figure*}

In the GHW Eq.(\ref{GHE-eq}), the norm kernel 
${\cal{N}}(\vec{\beta}_{1},\vec{\beta}_{2})$ accounts for the 
non-orthogonality of the  HFB states $| {\varphi} (\vec{\beta}) 
\rangle$. Therefore, the functions $f^{\sigma} (\vec{\beta})$ cannot be 
interpreted as probability amplitudes \cite{rs}. In order to obtain a 
probabilistic quantum mechanical interpretation, one introduces the 
collective wave functions 
\begin{equation} \label{cll-wfs-HW} 
G^{\sigma} (\vec{\beta}_{1}) =   \int_{{\cal{D}}_{\vec{\beta}_{2}}} d\vec{\beta}_{2} ~
{\cal{N}}^{\frac{1}{2}}(\vec{\beta}_{1},\vec{\beta}_{2}) 
f^{\sigma} (\vec{\beta}_{2})
\end{equation}
written in terms of the operational square root of the norm kernel 
\cite{rs,Sheikh21,2DGCM-q2q3-Gogny-1}. For the  state $| 
{\Psi}_{2D-GCM}^{\sigma} \rangle$ Eq.(\ref{GCM-WF}), the dynamical 
quadrupole and hexadecapole moments $Q_{\lambda 0,2D-GCM}^{\sigma}=  
\langle {\Psi}_{2D-GCM}^{\sigma} | \hat{Q}_{\lambda 0} | 
{\Psi}_{2D-GCM}^{\sigma} \rangle $ ($\lambda = 2,4)$  can be computed 
using the general expressions developed in previous works 
\cite{2DGCM-q2q3-Gogny-1,2DGCM-q2q3-Gogny-2, 
2DGCM-q2q3-Gogny-3,2DGCM-q2q3-Gogny-4,2DGCM-q2q3-Gogny-5}. The 
deformation parameters $\beta_{\lambda,2D-GCM}^{\sigma}$ are related to 
the moments  $Q_{\lambda 0,GCM}^{\sigma}$ according to 
Eq.(\ref{def-betas}). Let us also mention, that the overlaps of 
operators between different HFB states, required at the GCM level, have 
been computed using pfaffian techniques \cite{pfafian-1,pfafian-2}.

%
%
%

\section{Discussion of the results}
\label{results}


The results of the calculations for 
 $^{232-268}$Ra, $^{232-268}$Th, $^{232-268}$U 
and $^{232-268}$Pu are discussed in this section. First, mean-field results will be presented
in Sec.~\ref{MF_RESULTS}, while GCM results will be analyzed
in Sec.~\ref{GCM_RESULTS}.


\subsection{Mean-field calculations}
\label{MF_RESULTS}


The Gogny-D1S MFPESs obtained for the uranium isotopes $^{232-268}$U 
are depicted in Fig.\ref{mean-field-surfaces-U-1}, as illustrative 
examples. Similar results have been obtained for $^{232-268}$Ra, 
$^{232-268}$Th and $^{232-268}$Pu. Most of  the studied nuclei exhibit  
prolate  ground states. Up to around the mass number $A=248$, the 
ground state quadrupole deformation displays variations within the 
range  $0.24 \le \beta_{2} \le 0.28$. On the other hand, for larger 
mass numbers those values decrease almost linearly and HFB ground 
states with   $-0.12 \le \beta_{2} \le 0.0$ are predicted for 
$^{266,268}$Ra, $^{266,268}$Th, $^{266,268}$U and $^{268}$Pu.

As can be noted from Fig.\ref{mean-field-surfaces-U-1}, sizable static 
hexadecapole deformations  are predicted around $^{238}$U. This agrees 
well with the conclusions of Ref.~\cite{b4-RIHC}. For example, in our 
calculations we have obtained $\beta_{4}$ = 0.09, 0.13, 0.16 and  0.17 
for $^{238}$Ra, $^{238}$Th, $^{238}$U and $^{238}$Pu, respectively. 
With increasing mass number, the HFB  $\beta_{4}$ values decrease 
steadily up to  $\beta_{4}$ = 0 for $^{248}$Ra, $^{250}$Th, $^{252}$U 
and $^{254}$Pu. For larger mass numbers, as we move closer to the 
neutron shell closure $N = 184$ \cite{triaxial-example-6}, a region 
(i.e., the nuclei $^{250-264}$Ra, $^{252-264}$Th, $^{254-264}$U and 
$^{256-266}$Pu) with  $-0.04 \le \beta_{4} \le -0.02$ is predicted for 
the considered isotopic chains. Given the small negative $\beta_{4}$ 
values obtained for those neutron-rich isotopes, and the relatively 
flat potential energy surface in the $\beta_{4}$ direction, a 
beyond-mean-field analysis is required to examine the survival of the 
corresponding static hexadecapole deformations  once zero-point quantum 
fluctuations in the collective $(\beta_{2},\beta_{4})$-coordinates are 
taken into account.

\begin{figure*}
\includegraphics[width=0.95\textwidth]{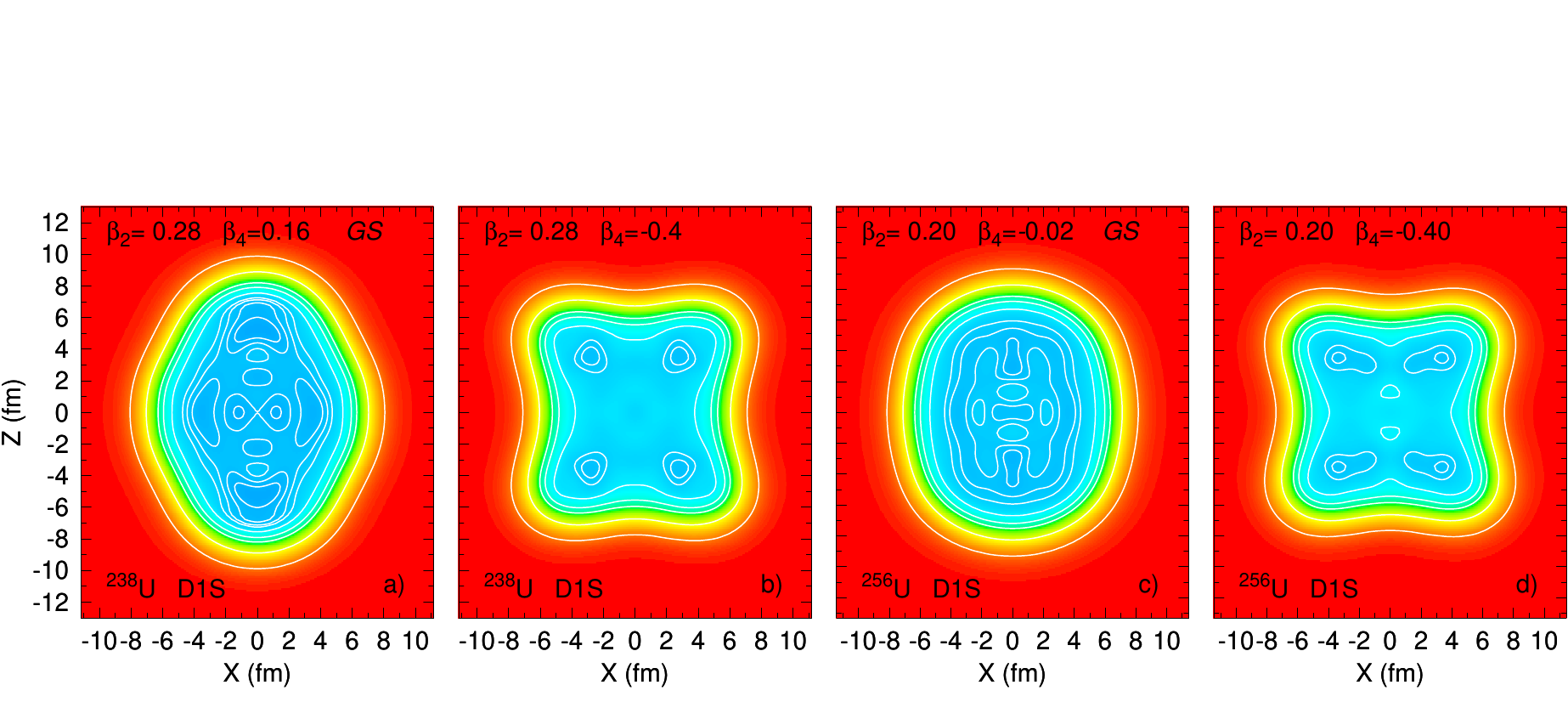}
\caption{(Color online) 
Mean-field density contour plots for the isotopes $^{238}$U [panels a) 
and b)] and $^{256}$U [panels c) and ()]. The ground state (GS) in 
$^{238}$U [panel a)] corresponds to $\beta_{2}$= 0.28 and $\beta_{4}$= 
0.16, while the GS in $^{256}$U [panel c)] corresponds to $\beta_{2}$= 
0.20 and $\beta_{4}$= -0.02. Typical square-like shapes corresponding 
to a large negative $\beta_{4}$= -0.4 deformation parameter are  
plotted in panels b) and d). Densities are in units of fm$^{-3}$ and 
contour lines at drawn at 0.01, 0.05, 0.10, 0.12, 0.14,  0.15, 0.16 and 0.17 fm$^{-3}$. Results 
have been obtained with the Gogny-D1S EDF.
}
\label{densities-mean-field-b2-b4} 
\end{figure*}

The HFB density contour plots  corresponding to the ground states 
$(\beta_{2}=0.28,\beta_{4}=0.16)$ and $(\beta_{2}=0.20,\beta_{4}=-0.02)$ of
$^{238}$U and $^{256}$U are plotted in panels (a) and (c)
of Fig.\ref{densities-mean-field-b2-b4}. In the case of $^{238}$U, one 
observes a typical diamond-like shape corresponding to positive $\beta_{4}$, while
an incipient square-like shape corresponding to a small, negative $\beta_{4}$
value is obtained for  $^{256}$U. Typical square-like shapes
corresponding to a large negative $\beta_{4}$= -0.4 value are 
illustrated for both isotopes in panels (b) and (d) of the figure. 

Coming back to the MFPESs displayed in Fig. 
\ref{mean-field-surfaces-U-1}, one realizes that for the lightest U 
isotopes, the excitation energy of oblate configurations is large and, 
therefore, their contribution to the ground state dynamics can be 
safely neglected. We have  checked this point by performing 2D-GCM 
calculations for the  isotopes $^{232-240}$U in which only 
configurations  with $\beta_{2} \ge 0$ have been included in the 2D-GCM 
state Eq.(\ref{GCM-WF}). However, the structural evolution of the 
MFPESs in Fig.\ref{mean-field-surfaces-U-1} also reveals that as we 
move to more neutron-rich sectors of the considered  isotopic chains, 
oblate configurations decrease their excitation energy and, at least 
for some of the considered nuclei, there is a pronounced competition 
between different intrinsic (quadrupole and/or hexadecapole) shapes. 
For instance, in the case of $^{260}$U and $^{264}$U the oblate 
configurations at $(\beta_{2}=-0.14,\beta_{4}=0.04)$  and 
$(\beta_{2}=-0.12,\beta_{4}=0.02)$ lie 1.50 and 0.38 MeV above the 
corresponding ground states. Moreover, the MFPESs obtained for the 
oblate-deformed isotopes  $^{266}$U and $^{268}$U exhibit a soft 
behavior along the quadrupole direction. Similar results, have been 
obtained for Ra, Th and Pu nuclei.

In Fig.\ref{mean-field-surfaces-U-1} we also observe that, for several 
of the studied nuclei (i.e., $^{232-256}$U), the corresponding MFPES 
displays an energy valley that, in the neighborhood of the absolute 
minimum, roughly follows a straight line with positive slope in the 
$(\beta_{2}, \beta_{4})$-plane for prolate deformations. This straight 
line corresponds to the direction labeled as A in the figure. Along 
this direction A, the hexadecapole deformation $\beta_{4}$ exhibits a 
linear behavior $\beta_{4}= a_{A} \beta_{2} + b_{A}$ as a function of  
$\beta_{2}$.  The line B, defines the direction perpendicular to A. 
Along the direction B, $\beta_{4}$ also exhibits a linear behavior 
$\beta_{4}= - (1/a_{A}) \beta_{2} + b_{B}$. For each nucleus, we have 
approximately fitted $a_{A}$, $b_{A}$ and $b_{B}$ using configurations 
around the absolute minimum of the MFPES. For example, in the case of 
$^{238}$U the directions A and B are parameterized as $\beta_{4}= 2.50 
\beta_{2} - 0.52$ and $\beta_{4}= -0.40 \beta_{2} + 0.29$, 
respectively.

\begin{figure}
\includegraphics[width=0.49\textwidth]{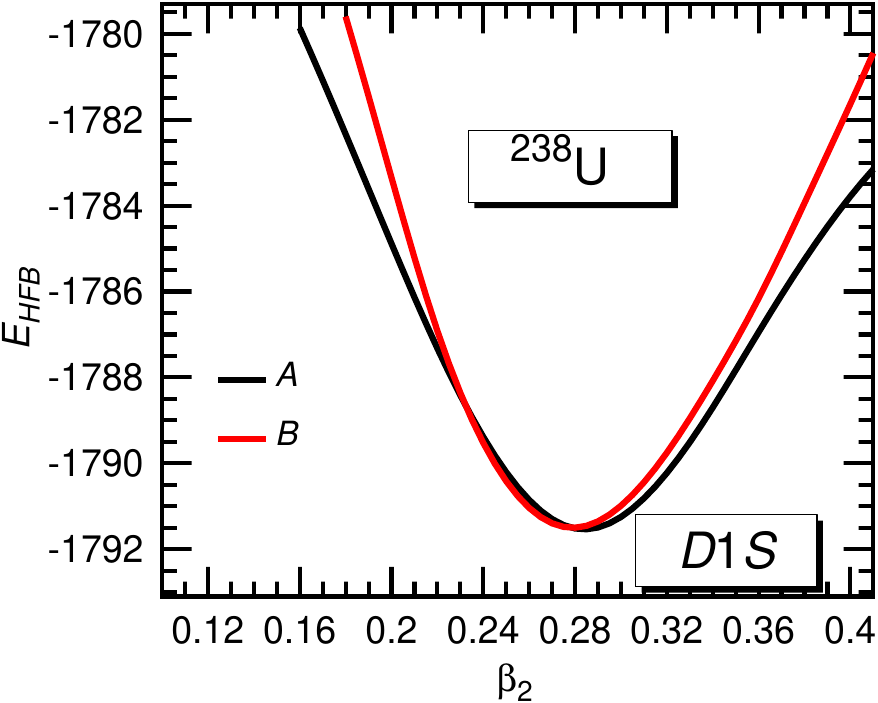}
\caption{(Color online) HFB energy, as a function of the deformation parameter 
$\beta_{2}$, along the directions A ($\beta_{4}= 2.50 \beta_{2} - 
0.52$) and B ($\beta_{4}= -0.40 \beta_{2} + 0.29$) for the nucleus 
$^{238}$U. Results have been obtained with the Gogny-D1S EDF. For more 
details, see the main text.
}
\label{pathsAB-Energy} 
\end{figure}

The energy $E_{HFB}$, as a function of the deformation parameter 
$\beta_{2}$, along the directions A  and B is depicted in 
Fig.\ref{pathsAB-Energy}  for  $^{238}$U. As can be seen, it exhibits a 
parabolic behavior in both cases. Thus, as indicated in the caption of 
Fig.\ref{mean-field-surfaces-U-1}, the directions A and B are nothing 
else than the principal axes of the two-dimensional parabola that 
approximates the HFB energy around the absolute minimum of the MFPES.

The tilted parabolic behavior of $E_{HFB}$ already described implies, 
that 1D-GCM calculations using  $\beta_{2}$ or $\beta_{4}$ as  single 
generating coordinates roughly explore the same configurations around 
the minimum of the MFPES \cite{large-scale-b4}. Thus, for several of 
the studied U but also Ra, Th and Pu isotopes, the quadrupole and 
hexadecapole degrees of freedom cannot be decoupled and full-fledged 
2D-GCM calculations must be carried out. As discussed below, a cheaper 
alternative to the 2D-GCM calculation is to perform 1D-GCM calculations 
along paths A and B discussed above. Note that, in particular, 
$^{238}$U belongs to this coupled regime. This is at variance with the 
rather weak quadrupole-octupole coupling, observed in previous studies 
\cite{2DGCM-q2q3-Gogny-1,2DGCM-q2q3-Gogny-2, 
2DGCM-q2q3-Gogny-3,2DGCM-q2q3-Gogny-4,2DGCM-q2q3-Gogny-5}. The 
exceptions in our calculations are neutron-rich isotopes, such as 
$^{258-268}$U, for which the directions A and B run parallel to the 
quadrupole and hexadecapole axes, respectively. In those cases, one 
might anticipate a weak coupling of the quadrupole and hexadecapole 
degrees of freedom.

\subsection{GCM calculations}
\label{GCM_RESULTS}

\begin{figure*}
\includegraphics[width=1.0\textwidth]{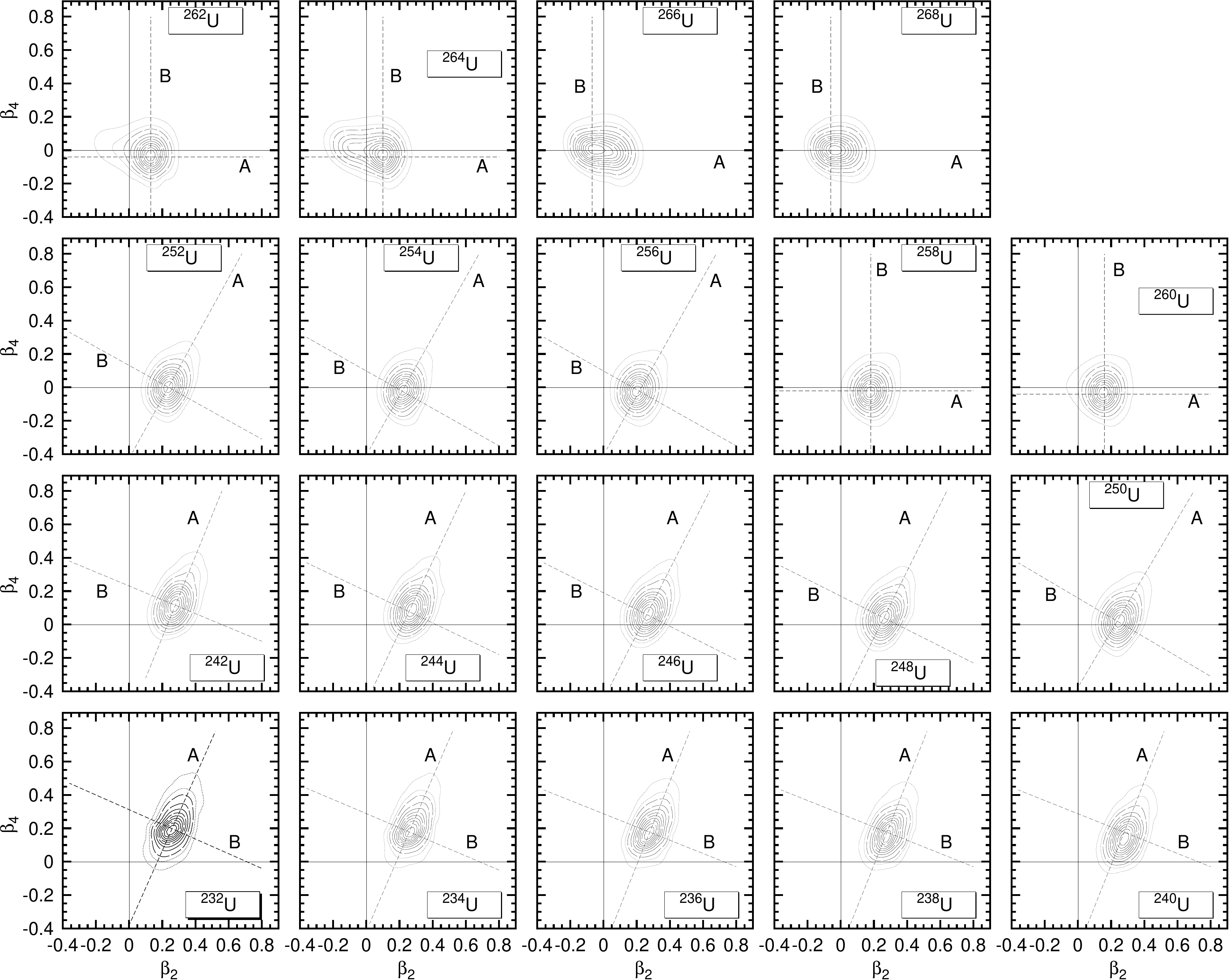}
\caption{Collective wave functions Eq.(\ref{cll-wfs-HW}) 
corresponding to the ground states of the nuclei $^{232-268}$U. The 
succession of solid, long dashed and short dashed contour lines starts 
at 90$\%$ of the maximum value up to 10$\%$ of it. The two dotted-line 
contours correspond to the tail of the amplitude (5$\%$ and 1$\%$ of 
the maximum value).  For each nucleus, the two perpendicular dotted lines A 
and B are drawn along the principal axes of the parabola that 
approximates the HFB energy around the absolute minimum of the MFPES 
and are the same as in Fig.\ref{mean-field-surfaces-U-1}. 
A vertical full line is drawn to signal the $\beta_{2}=0$
line whereas a full horizontal line is drawn to signal the $\beta_{4}=0$ line.
Results have 
been obtained with the Gogny-D1S EDF. For more details, see the main 
text.
}
\label{GS_CWF_U} 
\end{figure*}

The ground state collective  wave functions $G^{\sigma=1} 
(\vec{\beta})$ obtained for $^{232-268}$U are shown in Fig. 
\ref{GS_CWF_U}, as illustrative examples. It becomes apparent from the 
figure that for $^{232-256}$U, those  wave functions are well approximated
by a Gaussian with principal directions 
aligned along the directions A and B, already discussed in 
Sec.~\ref{MF_RESULTS}. On the other hand, for $^{258-268}$U 
they align parallel to the $\beta_{2}$ axis  and, therefore, in those 
isotopes the $(\beta_{2},\beta_{4})$-coupling is rather weak. This is 
similar to the situation encountered in previous dynamical studies 
(see, for example, \cite{2DGCM-q2q3-Gogny-5}) in which the 
quadrupole-octupole coupling has been shown to be rather weak and the 
corresponding ground state collective wave functions align along the 
octupole axis. Regardless of their particular alignment, the maxima of 
the collective wave functions in Fig.\ref{GS_CWF_U} correspond to 
$\beta_{2}$ and $\beta_{4}$ values close to the absolute minima of the 
MFPESs.

For the other isotopic chains the nuclei signaling the transition from
coupled to uncoupled $(\beta_{2},\beta_{4})$ regime also correspond to
neutron number 166-168. 

The previous results agree well with the structural evolution of the 
MFPESs and corroborate, from a dynamical point of view, the transition 
from a $(\beta_{2},\beta_{4})$-coupled to a 
$(\beta_{2},\beta_{4})$-decoupled regime in the 2D-GCM ground states of 
the studied nuclei with increasing mass number. Furthermore, they 
corroborate the relevance \cite{large-scale-b4} of collective 
deformations along the A and B directions as alternative generating  
coordinates  in GCM calculations for the considered Ra, Th, U and Pu 
nuclei. With this in mind, we have also performed 1D-GCM calculations 
along those directions for all the studied isotopes. 

\begin{figure*}
\includegraphics[width=1.00\textwidth]{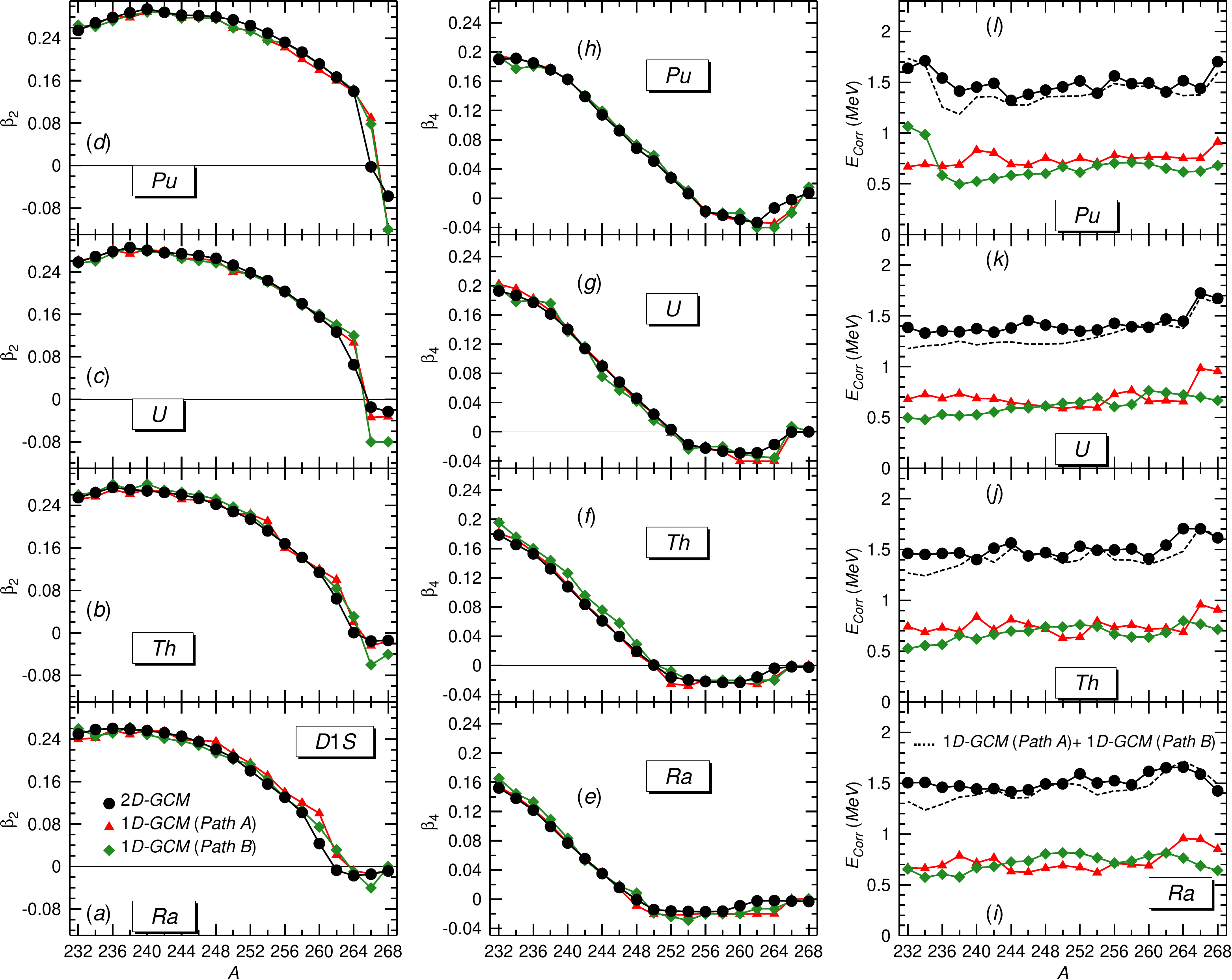}
\caption{(Color online) 2D-GCM ground state quadrupole $\beta_{2}$ 
[panels (a)-(d)] and hexadecapole $\beta_{4}$ [panels (e)-(h)] 
deformation parameters for $^{232-268}$Ra, $^{232-268}$Th, 
$^{232-268}$U and $^{232-268}$Pu. The 2D-GCM correlation energies 
$E_{Corr,2D-GCM}$ obtained for those nuclei are depicted in panels (i)-(l). 
Quadrupole and hexadecapole deformations as well as correlation 
energies obtained in 1D-GCM calculations along the directions A and B 
are also included in the plots. The dotted lines in panels (i)-(l) 
correspond to the sum of 1D-GCM correlation energies along the 
directions A and B. Results have been obtained with the Gogny-D1S EDF. 
For more details, see the main text. 
}
\label{2DGCM-b2-b4} 
\end{figure*}

The ground state quadrupole $\beta_{2,2D-GCM}^{\sigma=1}$ and 
hexadecapole $\beta_{4,2D-GCM}^{\sigma=1}$ deformation parameters 
obtained for $^{232-268}$Ra, $^{232-268}$Th, $^{232-268}$U and 
$^{232-268}$Pu are depicted in panels (a)-(d) and (e)-(h) of 
Fig.\ref{2DGCM-b2-b4}, respectively. Their values are rather close to 
the ones obtained at the HFB level. In particular, the 
$\beta_{2,GCM}^{\sigma=1}$ deformations slightly increase up to 0.26, 
0.27, 0.29 and 0.29 in $^{236}$Ra, $^{236}$Th, $^{238}$U and 
$^{240}$Pu, respectively, while for larger mass numbers they decrease 
until 2D-GCM ground states with $-0.06 \le \beta_{2,2D-GCM}^{\sigma=1} 
\le 0.00$ are obtained for  $^{262-268}$Ra, $^{264-268}$Th, 
$^{266-268}$U and $^{266-268}$Pu.

As can be seen from panels (e)-(h), sizable dynamical hexadecapole 
$\beta_{4,GCM}^{\sigma=1}$ values are predicted around $^{238}$U 
\cite{b4-RIHC}. For example, we have obtained 
$\beta_{4,GCM}^{\sigma=1}$ = 0.10, 0.13, 0.16 and  0.18 for $^{238}$Ra, 
$^{238}$Th, $^{238}$U and $^{238}$Pu, respectively. Furthermore, our 
2D-GCM calculations corroborate the dynamical survival of  regions 
(i.e., the nuclei $^{250-260}$Ra, $^{252-262}$Th, $^{254-264}$U and 
$^{256-264}$Pu) with small negative hexadecapole deformations $- 0.034 
\le \beta_{4,GCM}^{\sigma=1} \le -0.014$ in each of the considered 
isotopic chains, just below the neutron magic number  $N = 184$ 
\cite{large-scale-b4,polar-gap-1,polar-gap-2}.

In panels (i)-(l) of Fig.~\ref{2DGCM-b2-b4}, we have plotted the 2D-GCM 
correlation energies defined as the difference 
\begin{equation} 
E_{Corr,2D-GCM} = E_{HFB,gs} - E^{\sigma=1}
\end{equation}
between the HFB $E_{HFB,gs}$ and 2D-GCM $E^{\sigma=1}$ ground state 
energies. For the studied nuclei, $1.42$ MeV $\le$ $E_{Corr,2D-GCM}$ 
$\le$ $1.72$ MeV. The range of variation of the correlation energies 
(0.3 MeV) is comparable to typical values of the rms for the binding 
energy in Gogny-like mass tables \cite{gogny-d1m,gogny-d1mstar}. 
Therefore,  the correlation energies $E_{Corr,2D-GCM}$ should be taken 
into account in future fitting protocols of the Gogny-EDF.

A more quantitative measure of the role of hexadecapole deformation and 
its coupling to the quadrupole degree of freedom in the ground state 
dynamics, can be obtained from the  comparison  with   correlation 
energies $E_{Corr,1D-GCM}^{Q2}$ resulting from 1D-GCM calculations 
based on the ansatz
\begin{equation} \label{GCM-WF-Q2}
| {\Psi}^{\sigma} \rangle = \int_{{\cal{D}}_{2}} d\beta_{2} f^{\sigma} (\beta_{2}) | {\varphi} (\beta_{2}) \rangle
\end{equation} 
in which, the basis states $| {\varphi} (\beta_{2}) \rangle$ are 
obtained via $\beta_{2}$-constrained HFB calculations in the domain 
${{\cal{D}}_{2}}= \Big\{\beta_{2} \in [-0.4,1.2] \Big\}$ with the step 
size $\delta \beta_{2}= 0.01$. For the studied nuclei, we have obtained 
$0.45$ MeV $\le$ $E_{Corr,1D-GCM}^{Q2}$ $\le$ $0.98$ MeV. Thus, in 
going from the (quadrupole) 1D- to the (quadrupole-hexadecapole) 
2D-GCM, we gain an additional correlation energy  $0.74$ MeV $\le$ 
$\delta E_{Corr}$ $\le$ $0.97$ MeV similar to the quadrupole 
correlation energy itself. These results, and the ones obtained in 
previous studies \cite{2DGCM-q2q3-Gogny-1,2DGCM-q2q3-Gogny-2, 
2DGCM-q2q3-Gogny-3,2DGCM-q2q3-Gogny-4,2DGCM-q2q3-Gogny-5,large-scale-b4}, 
suggest a slow convergence of the nuclear correlation energy with 
respect to the (even and/or odd) shape multipole moments included 
within the GCM framework and deserve further attention in future work.

\begin{figure*}
\includegraphics[width=1.0\textwidth]{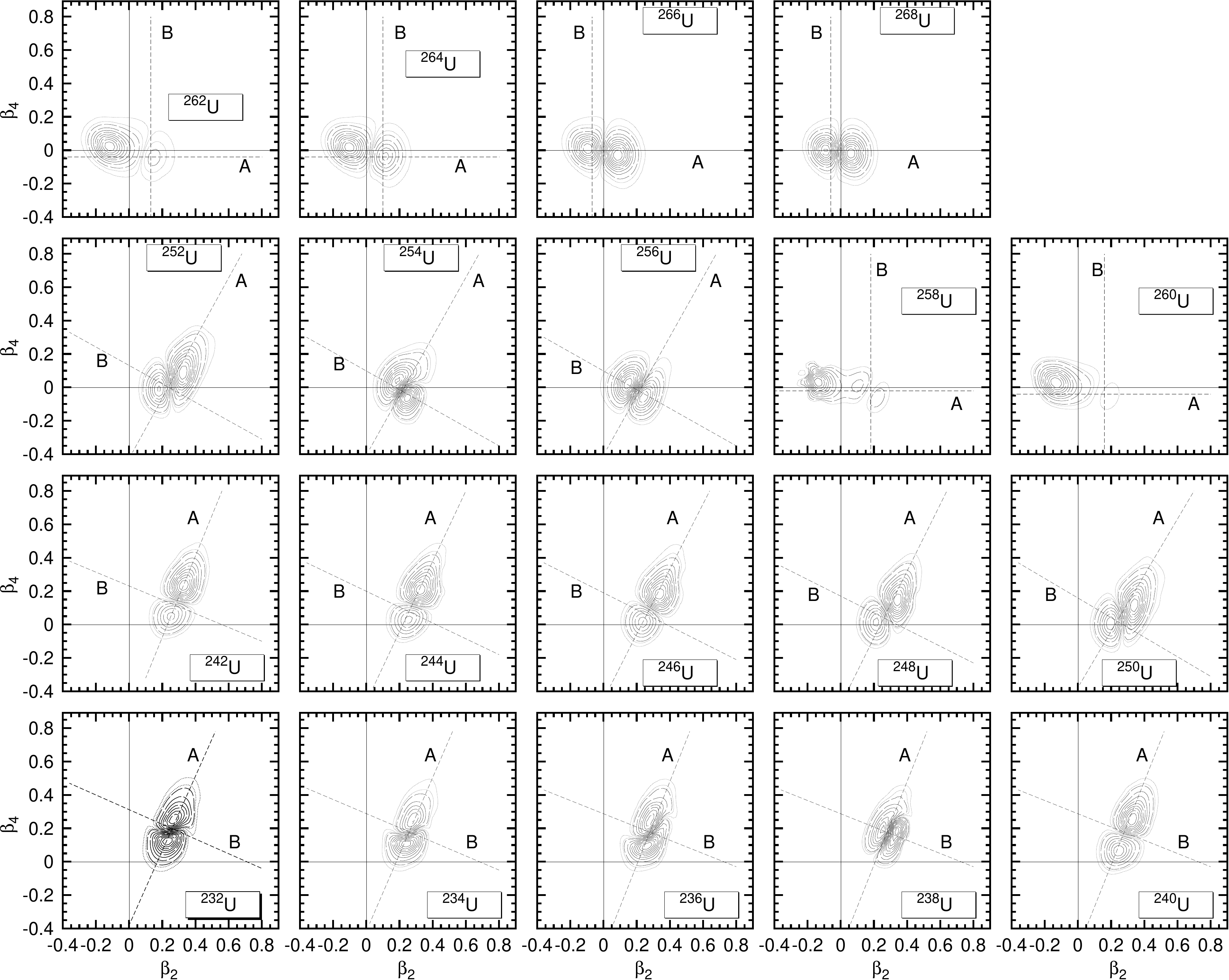}
\caption{The same as in Fig.~\ref{GS_CWF_U} but for the first 
excited 2D-GCM states of the nuclei $^{232-268}$U. 
}
\label{FEXC_CWF_U} 
\end{figure*}

Given the relevance of the directions A and B, we have also carried out 
1D-GCM calculations along those directions. Within this context, we 
have considered the ansatz 
\begin{equation} \label{GCM-WF-PathA} | {\Psi}_{A}^{\sigma} \rangle = 
\int_{{\cal{D}}_{A}} d\beta_{2} f_{A}^{\sigma} (\beta_{2}) | {\varphi} 
(\beta_{2}) \rangle \end{equation} 
where the integration domain reads 
${{\cal{D}}_{A}}= \Big\{\beta_{2} \in [-0.4,1.2]; \beta_{4}= a_{A} 
\beta_{2} + b_{A} \Big\}$. Another set of calculations resorted to a 
similar ansatz 
\begin{equation} \label{GCM-WF-PathB}
| {\Psi}_{B}^{\sigma} \rangle = \int_{{\cal{D}}_{B}} d\beta_{2} f_{B}^{\sigma} (\beta_{2}) | {\varphi} (\beta_{2}) \rangle
\end{equation}
with the domain ${{\cal{D}}_{B}}= \Big\{\beta_{2} \in [-0.4,1.2]; 
\beta_{4}= -(1/a_{A}) \beta_{2} + b_{B} \Big\}$. In both sets, we have 
employed the step $\delta \beta_{2}= 0.01$. For each nucleus, the 
parameters $a_{A}$, $b_{A}$ and $b_{B}$ have been fitted along the 
lines already discussed in Sec.~\ref{MF_RESULTS}. As can be seen from 
panels (a)-(d) and (e)-(h) of Fig.\ref{2DGCM-b2-b4}, the quadrupole and 
hexadecapole deformations corresponding to the  1D- and 2D-GCM ground 
states $| {\Psi}_{A}^{\sigma=1} \rangle$, $| {\Psi}_{B}^{\sigma=1} 
\rangle$ and $| {\Psi}_{2D-GCM}^{\sigma=1} \rangle$ are rather similar. 

\begin{figure*}
\includegraphics[width=1.0\textwidth]{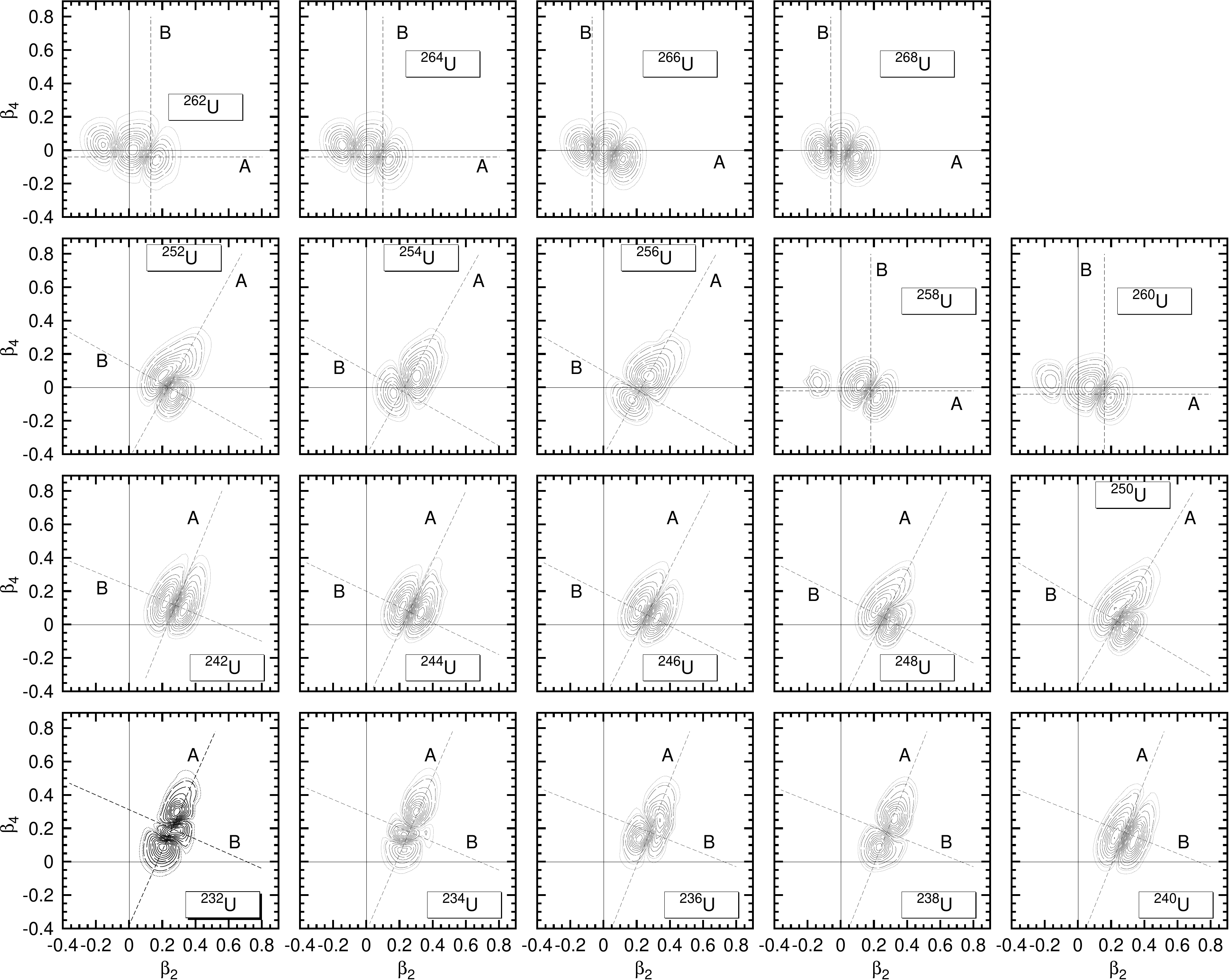}
\caption{The same as in Fig.~\ref{GS_CWF_U} but for the second 
excited 2D-GCM states of the nuclei $^{232-268}$U. 
}
\label{SECONDEXC_CWF_U} 
\end{figure*} 

In panels (i)-(l) of Fig.\ref{2DGCM-b2-b4}, we have also included the 
1D-GCM correlation energies $E_{Corr,1D-GCM}^{A}$ and 
$E_{Corr,1D-GCM}^{B}$. In most cases, those energies are rather similar 
and, as expected, the A path results compare well with the 
corresponding quadrupole $E_{Corr,1D-GCM}^{Q2}$ values.  Moreover, 
the sum $E_{Corr,1D-GCM}^{A}+ E_{Corr,1D-GCM}^{B}$ approximates 
$E_{Corr,2D-GCM}$ reasonably well. Let us also stress, that in going 
from 1D-GCM calculations along the directions A and B to the 2D-GCM 
ones we gain the additional correlation energies $0.74$ MeV $\le$ 
$\delta E_{Corr}$ $\le$  $0.81$ MeV and $0.66$ MeV $\le$ $\delta 
E_{Corr}$ $\le$  $0.94$ MeV, respectively. Once more, these results 
point towards the key role of the $(\beta_{2},\beta_{4})$ dynamical 
zero-point fluctuations in the ground states of the considered nuclei.

The collective  wave functions $G^{\sigma=2} (\vec{\beta})$ 
corresponding to the  first excited states in $^{232-268}$U are shown 
in Fig.\ref{FEXC_CWF_U}. For the isotopes $^{232-236}$U and 
$^{240-252}$U ($^{238}$U and $^{254,256}$U) they correspond to phonons 
aligned along the tilted direction A (B). In such phonons the 
quadrupole and hexadecapole degrees of freedom are coupled. On the 
other hand, the wave functions $G^{\sigma=2} (\vec{\beta})$ in 
$^{258-268}$U align parallel to the quadrupole axis. For both 
$^{258-262}$U, the collective amplitude  mainly concentrates on the 
oblate minimum and corresponds to a shape isomeric excited state, while 
for  $^{264-268}$U the first excited state slowly evolves from the 
oblate configuration to a pure two-phonon quadrupole vibration.

The analysis of the collective amplitudes $G^{\sigma=3} 
(\vec{\beta})$, associated with the second excited states, shown in 
Fig.\ref{SECONDEXC_CWF_U}, reveals a quite rich casuistic. The amplitudes
in the isotopes $^{232-236}$U correspond to the second vibration phonon
along direction A. In the case of $^{238}$U the first excited state corresponds to a phonon along the 
direction B, while the second excited state represents a phonon along 
the direction A. For the isotopes $^{240-252}$U the second excited state
corresponds to a phonon along the B direction. However, for $^{254-260}$U
the second excited state corresponds to a single phonon along the A direction,
whereas for $^{262-268}$U two phonon excitations along the A direction
are observed. 
As a general conclusion, not only for the ground  but 
also for the first and second excited states of the studied nuclei, our 
calculations predict a  transition from a 
$(\beta_{2},\beta_{4})$-coupled to a $(\beta_{2},\beta_{4})$-decoupled 
regime with increasing mass number. 

The analysis of all the collective amplitudes discussed above reveal 
that while the ground state can be well described by two 1D-GCM calculations
along the A and B directions, the situation becomes more intricate as
one considers higher excited states (consider, for instance, $^{250-252}$~U) 
and there are some specific cases where the 2D-GCM calculation is required.
Unfortunately, there is not a clear trend with neutron number and our
recommendation is to perform the full fledged 2D-GCM calculation when
the properties of excited states are required.

\begin{figure*}
\includegraphics[width=1.00\textwidth]{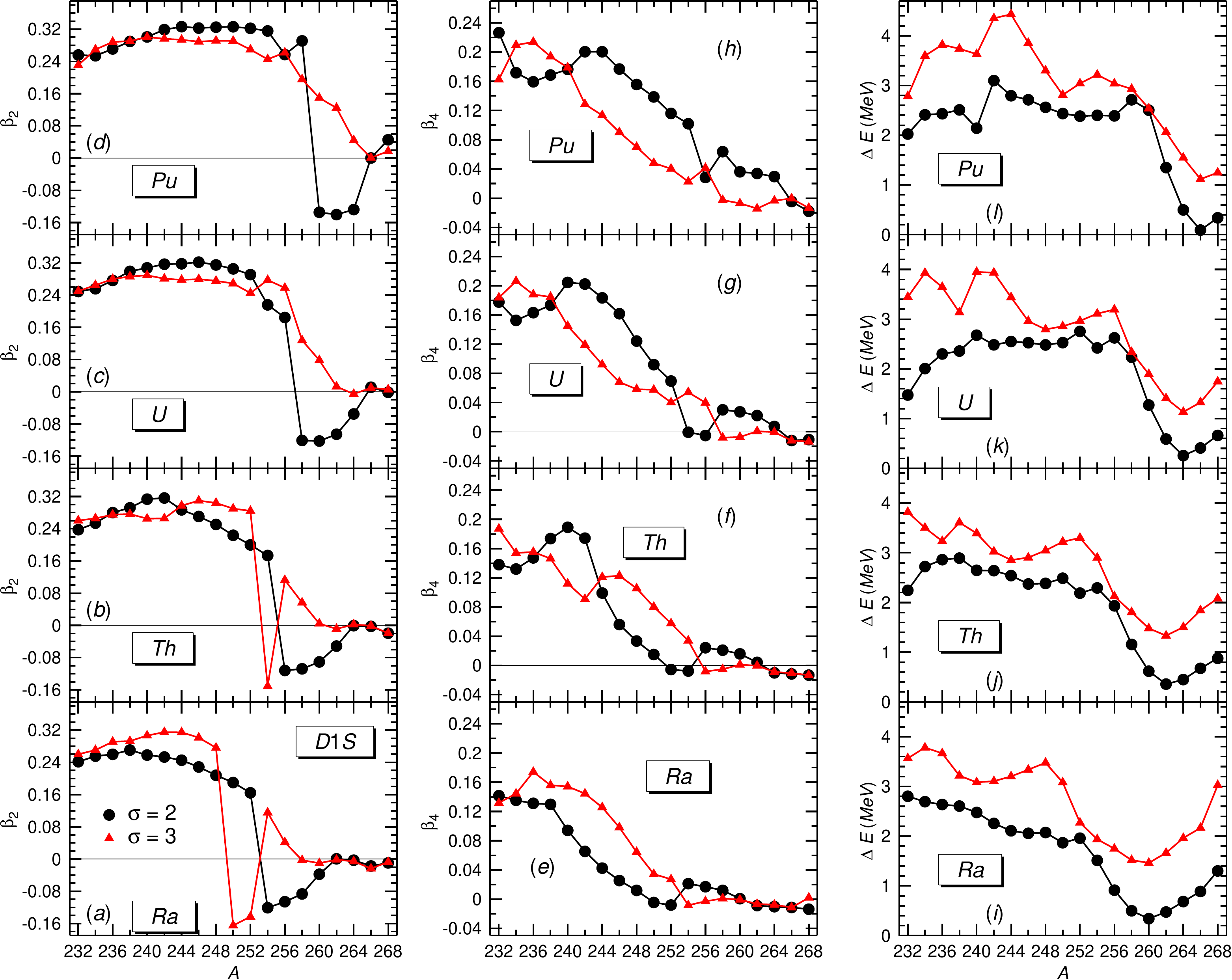}
\caption{(Color online) 2D-GCM first ($\sigma =2$) and second ($\sigma 
=3$) excited states quadrupole $\beta_{2}$ [panels (a)-(d)] and 
hexadecapole $\beta_{4}$ [panels (e)-(h)] deformation parameters for 
$^{232-268}$Ra, $^{232-268}$Th, $^{232-268}$U and $^{232-268}$Pu. The 
excitation energies $\Delta E^{ \sigma=2}$ and $\Delta E^{ \sigma=3}$ 
corresponding to those states are depicted in panels (i)-(l). Results 
have been obtained with the Gogny-D1S EDF. For more details, see the 
main text. 
}
\label{2DGCM-b2-b4-FEXC} 
\end{figure*}

The 2D-GCM quadrupole deformations $\beta_{2,2D-GCM}^{\sigma=2}$ and 
$\beta_{2,2D-GCM}^{\sigma=3}$ obtained for  $^{232-268}$Ra, 
$^{232-268}$Th, $^{232-268}$U and $^{232-268}$Pu are plotted in panels 
(a)-(d) of Fig.\ref{2DGCM-b2-b4-FEXC}.  The first excited states of  
$^{232-252}$Ra, $^{232-254}$Th, $^{232-256}$U and $^{232-258}$Pu are 
prolate deformed. At variance with their ground states (see, 
Fig.\ref{2DGCM-b2-b4}), the first excited states of $^{254-260}$Ra, 
$^{256-262}$Th, $^{258-264}$U  and $^{260-264}$Pu exhibit oblate 
deformations within the range $ -0.14 \le \beta_{2,2D-GCM}^{\sigma=2} 
\le -0.04$.  As already discussed in Sec.\ref{MF_RESULTS} (see, 
Fig~\ref{mean-field-surfaces-U-1}), for the more neutron-rich sectors 
in each of the studied isotopic chains, the MFPESs exhibit soft 
features as well as a more pronounced competition between low-lying 
configurations based on different intrinsic shapes. From a dynamical 
point of view, both the 2D-GCM shape transitions in $^{254-260}$Ra, 
$^{256-262}$Th, $^{258-264}$U  and $^{260-264}$Pu and the 
$\beta_{2,2D-GCM}^{\sigma=2}$ values obtained for heavier isotopes, 
reflect this enhanced shape coexistence and result from the structure 
of the corresponding collective wave functions $G^{\sigma=2} 
(\vec{\beta})$ in Fig.\ref{FEXC_CWF_U}. Similar arguments hold for the 
$\beta_{2,2D-GCM}^{\sigma=3}$ deformations resulting from the structure 
of the wave functions $G^{\sigma=3} (\vec{\beta})$ in 
Fig.\ref{SECONDEXC_CWF_U}. Note that, the sudden jump to
$\beta_{2,2D-GCM}^{\sigma=3} < 0$  values observed in 
$^{250-252}$Ra and $^{254}$Th results from the concentration 
of a significant portion of the corresponding collective  
$G^{\sigma=3} (\vec{\beta})$ strengths on the oblate side.

The  hexadecapole deformations $\beta_{4,2D-GCM}^{\sigma=2}$ and 
$\beta_{4,2D-GCM}^{\sigma=3}$ corresponding to  the first and second 
excited states in $^{232-268}$Ra, $^{232-268}$Th, $^{232-268}$U and 
$^{232-268}$Pu are shown in panels (e)-(h) of 
Fig.\ref{2DGCM-b2-b4-FEXC}. In our calculations, we have obtained 
$\beta_{4,GCM}^{\sigma=2}$ = 0.09, 0,17, 0.17 and  0.17 
($\beta_{4,GCM}^{\sigma=3}$ = 0.16, 0.14, 0.18 and 0.19) for the first 
(second) excited states in $^{238}$Ra, $^{238}$Th, $^{238}$U and 
$^{238}$Pu. Therefore, not only for the ground but also for the first 
two excited states of nuclei around $^{238}$U hexadecapole deformations 
play a significant role \cite{b4-RIHC}. For Ra isotopes, the 
$\beta_{4,2D-GCM}^{\sigma=2}$ deformations exhibit a smooth decrease up 
to the mass number $A = 252$. On the other hand, in the case of  Th, U 
and Pu nuclei, those  $\beta_{4,2D-GCM}^{\sigma=2}$ values display a 
steady decrease after reaching a maximum for $A = 240,242$. At variance 
with  their ground states, the first excited states of $^{254-260}$Ra, 
$^{256-262}$Th, $^{258-264}$U  and $^{260-264}$Pu correspond to 
diamond-like shapes.  Note that these are precisely the same nuclei, 
for which quadrupole transitions are observed in panels (a)-(d) of the 
figure.  On the other hand, for heavier isotopes weakly deformed 
square-like shapes are predicted. Besides their overall decreasing 
trend, the deformations $\beta_{4,2D-GCM}^{\sigma=3}$ also exhibit a 
transition to a weakly hexadecapole deformed regime around $A = 
254-258$. We stress, once more, that these features reflect the 
enhanced shape coexistence in the more neutron-rich sectors of the 
studied isotopic chains.

Finally, in panels (i)-(l) of Fig.~\ref{2DGCM-b2-b4-FEXC} we have 
depicted the excitation energies $\Delta E^{\sigma = 2}$ and $\Delta 
E^{\sigma = 3}$. The most striking feature observed in those panels is 
the pronounced reduction of the excitation energies  for the most 
neutron-rich Ra, Th, U and Pu isotopes with an almost parabolic 
behavior around their global minima for $^{260}$Ra, $^{262}$Th, 
$^{264}$U and $^{266}$Pu. In particular, we have obtained $\Delta 
E^{\sigma = 2}$ = 340, 360, 250, 91 KeV and $\Delta E^{\sigma = 3}$ = 
1.46, 1.33, 1.13 and 1.11 MeV. It is precisely the enhanced shape 
coexistence around those nuclei, what brings the excitation energies 
$\Delta E^{\sigma = 2}$ and $\Delta E^{\sigma = 3}$ to such low values 
as compared with lighter nuclei.

%
%

\section{Conclusions}
\label{conclusions}

In this work, we have addressed the emergence and stability of static 
hexadecapole deformation effects as
well as the impact of the dynamical quadrupole-hexadecapole 
configuration mixing  on a selected set of radium, thorium, uranium and plutonium
isotopes covering the mass numbers $ 232 \le A \le 268$. To this end, we have 
resorted to the $(\beta_{2},\beta_{4}$)-constrained Gogny-D1S HFB 
approximation, followed  by beyond-mean-field 2D-GCM calculations with
the deformation parameters $\beta_{2}$ and 
$\beta_{4}$ as generating coordinates.

From the results of both HFB and 2D-GCM calculations, we conclude that 
diamond-like shapes play a prominent role around $^{238}$U. In 
particular, sizable static and dynamic hexadecapole deformations are 
found not only for ground but also for the first two excited states of 
nuclei in the neighborhood of  $^{238}$U. This agrees well with the 
conclusions extracted from the analysis of hydrodynamic simulations of 
the quark-gluon plasma at the BNL Relativistic Heavy Ion Collider 
(RHIC) \cite{b4-RIHC}. In agreement with a previous HFB study 
\cite{large-scale-b4} and the  polar gap model 
\cite{polar-gap-1,polar-gap-2}, for each of the studied isotopic 
chains, our results corroborate the dynamical survival of a region 
characterized by square-like shapes (with associated weak hexadecapole 
deformations) just below the neutron magic number $N =184$. 

We have carried out a detailed analysis of the collective wave 
functions associated with both ground and excited states. We conclude 
that, with increasing mass number, a transition from a regime in which 
the quadrupole and hexadecapole degrees of freedom are interwoven  to  
a regime in which they are decoupled, takes place for the studied 
isotopic chains. This is accompanied by an enhanced quadrupole shape 
coexistence in the more neutron-rich sectors of those chains. This 
enhanced quadrupole shape coexistence, reflected in the structure of 
the corresponding collective wave functions, leads to low values of the 
excitation energies for $^{260}$Ra, $^{262}$Th, $^{264}$U and 
$^{266}$Pu.

We have obtained 2D-GCM ground state correlation energies which are of 
the same order of magnitude as the rms for the binding energy in 
Gogny-like mass tables \cite{gogny-d1m,gogny-d1mstar}. Therefore,  
those ground state correlation energies including the hexadecapole 
degree of freedom should also be considered in future parametrizations 
of the  Gogny-EDF. From the comparison of the 2D-GCM and  (quadrupole) 
1D-GCM correlation energies we conclude, that the inclusion of 
hexadecapole deformation in the ground state dynamics leads to an 
additional (correlation) energy gain which is comparable to the 
quadrupole correlation energy itself. Similar conclusions can be 
extracted from the comparison of 2D-GCM correlation energies with the 
ones obtained in 1D-GCM calculations based on generating coordinates 
essentially determined by the principal axes of the parabola 
approximating the energy around the HFB equilibrium configuration. 
Furthermore, the results of this and previous studies  
\cite{2DGCM-q2q3-Gogny-1,2DGCM-q2q3-Gogny-2, 
2DGCM-q2q3-Gogny-3,2DGCM-q2q3-Gogny-4,2DGCM-q2q3-Gogny-5,large-scale-b4} 
point towards a slow convergence of the correlation energy with respect 
to the (even and/or odd) shape multipole moments included within the 
GCM ansatz, and deserve attention in future works. 

\begin{acknowledgments}
The work of LMR is supported by Spanish Agencia Estatal de Investigacion 
(AEI) of the Ministry of Science and Innovation under Grant No. 
PID2021-127890NB-I00. The work of R. Rodr\'{\i}guez-Guzm\'an was partially 
supported through the grant PID2022-136228NB-C22 funded by 
MCIN/AEI/10.13039/501100011033/FEDER, UE and "ERDF A way of making 
Europe".
\end{acknowledgments}

\end{document}